\documentclass[conference]{IEEEtran}
\IEEEoverridecommandlockouts
\usepackage{amsmath,amssymb,amsfonts}
\usepackage{graphicx}
\usepackage{textcomp}\usepackage{booktabs}
\usepackage{pgfplots}
\pgfplotsset{compat=1.18}
\usepackage{xcolor}
\usepackage{graphicx}
\usepackage{booktabs}
\usepackage{longtable}
\usepackage{balance}
\usepackage{float}
\usepackage{pdflscape}
\usepackage{amssymb}
\usepackage{pifont}
\usepackage{listings}
\usepackage[utf8]{inputenc}
\usepackage[algo2e,ruled]{algorithm2e}
\usepackage{ifthen} 
\usepackage{array}
\usepackage{makecell}
\usepackage{booktabs}
\usepackage{flushend}
\usepackage{multicol}
\usepackage{amsmath,array,graphicx}
\usepackage[numbers,sort&compress]{natbib}
\usepackage{soul}
\usepackage{mathtools}
\usepackage{balance}
\usepackage[utf8]{inputenc}
\usepackage{mathtools}
\usepackage{comment}
\usepackage{listings}
\usepackage[utf8]{inputenc}
\usepackage{algorithm} 
\usepackage{algorithmic} 
\usepackage[algo2e,ruled]{algorithm2e}
\usepackage{ifthen} 
\usepackage[colorlinks=true, citecolor=blue]{hyperref}
\def\BibTeX{{\rm B\kern-.05em{\sc i\kern-.025em b}\kern-.08em
    T\kern-.1667em\lower.7ex\hbox{E}\kern-.125emX}}

\begin{document}

\title{ Structural Analysis of Cryptographic Sequences using Stringology-Based Fingerprinting\\
}
\author{ 
\IEEEauthorblockN{Victor Kebande\IEEEauthorrefmark{1}\IEEEauthorrefmark{2}}
\IEEEauthorblockA{\IEEEauthorrefmark{1}University of Colorado Denver, CO USA\\
\IEEEauthorrefmark{2}ATLAS Institute, University of Colorado 
Boulder, Colorado, USA\\
Email: victor.kebande@ucdenver.edu, victor.kebande@colorado.edu}
} 

\maketitle

\begin{abstract}

Cryptographic primitives such as stream ciphers, Pseudorandom Number Generators (PRNGs), and block cipher modes produce sequences that are designed to be statistically indistinguishable from random data. As a result, the traditional evaluation techniques therefore rely primarily on statistical randomness tests to assess the quality of generated sequences. While these tests verify global statistical properties, they do not address whether structural characteristics of sequences can reveal information about the underlying generator. In this paper, we introduce a \textit{stringology-based fingerprinting, (SBF)} framework for the structural analysis of cryptographic sequences. The proposed \textit{ SBF} framework  interprets cryptographic outputs as symbolic strings and applies pattern-based feature extraction to capture structural statistics such as substring frequency distributions, recurrence patterns, and entropy characteristics. These structural features are aggregated into fingerprint vectors that characterize sequence generators. The experimental evaluation is conducted using datasets composed of Cipher-Generated Sequences (CGS) and Uniformly Random Sequences (URS). The results demonstrate that stringology-based pattern analysis can reveal measurable structural signatures across different sequence sources. Although these signals do not imply practical cryptographic weaknesses, they provide an additional analytical perspective for evaluating the structural behavior of cryptographic generators. 


\end{abstract}


\section{Introduction}

Cryptographic systems are the cornerstone of modern digital communication infrastructures.  Many cryptographic primitives are able to produce sequential outputs like keystreams, Pseudorandom Number Generators (PRNGs) ciphertext blocks, and pseudorandom numbers that must appear to be statistically indistinguishable from random data, for purposes of ensuring  strong security guarantees \cite{micciancio2018bit}. As a result, evaluating the statistical properties of cryptographic sequences has long been an essential component of cryptographic design and analysis.

The traditional cryptographic evaluation techniques that uses   statistical randomness test suites such as the NIST Statistical Test Suite (STS), TestU01, etc have been seen to be effective \cite{kebande2023extended, zubkov2019testing, sleem2020testu01}. This, is owing to the fact that these frameworks evaluate global statistical properties including like distributions, run lengths, entropy, and correlation measures.

Even though these tests provide good conclusions on the randomness of the generated cryptographic sequences, they are only designed to detect significant deviations from ideal randomness as oposed to characterization of  the structural behavior of the underlying cryptographic generator. While the randomness test may pass, cryptographic sequences are still produced by deterministic algorithms composed of structured internal operations such as modular addition, bit rotations, substitutions, and XOR transformations \cite{blackledge2008multi}. These deterministic transformations may introduce subtle structural characteristics in generated sequences that are difficult to detect using purely statistical methods. 

From a structural perspective, these sequences can be interpreted as symbolic strings whose internal patterns may reflect the behavior of the algorithm that generated them. Stringology, which is  the study of algorithms for processing and analyzing strings \cite{watson2012correctness}, provides a natural framework for examining such structural properties. 


Motivated by this perspective, this paper introduces a \textit{stringology-based fingerprinting framework, (SBF)} for analyzing cryptographic sequences. The \textit{SBF}  extracts structural pattern statistics from sequences and then uses these statistics to construct fingerprint representations that characterize the behavior of sequence generators. These fingerprints capture pattern-based features such as substring frequency distributions, pattern recurrence statistics, and entropy-based measures that describe structural properties of the generated sequences. 

To evaluate the proposed framework, experiments are conducted using datasets composed of cipher-generated sequences and sequences drawn from a uniform random distribution. Pattern statistics are extracted using stringology-based feature analysis across multiple pattern lengths, and the resulting structural profiles are compared in order to identify measurable differences between sequence sources. 


The main contributions of this work can be summarized as follows: 
\begin{itemize} 
\item Introduce a \textit{SBF} framework for structural analysis of cryptographic sequences.
\item Propose a pattern-based fingerprinting representation that characterizes sequence generators using substring statistics. 

\item Demonstrate through experimental evaluation that stringology-based pattern analysis can reveal measurable structural differences between cryptographic and uniformly random sequences.

\end{itemize}

The remainder of this paper is organized as follows. Section~II presents background concepts on cryptographic sequence generators and stringology-based pattern analysis. Section~III reviews related work in cryptographic sequence evaluation. Section~IV describes the proposed stringology-based fingerprinting framework. Section~V presents the experimental evaluation. Section~VI reports the experimental results and analysis. Section~VII gives a discussion and implications of the findings, and Section~VIII concludes the paper and mentions a future work.

\section{Background}

\subsection{Cryptographic Sequence Generators}

Many cryptographic systems produce sequential outputs that are
intended to behave as pseudorandom sequences. Examples include
keystreams generated by stream ciphers, outputs from pseudorandom
number generators (PRNGs) \cite{kietzmann2021guideline}, and ciphertext streams produced by
block cipher modes such as CTR \cite{aljohani2019performance}. Formally, a Cryptographic Sequence
Generator \textit{CSG}  can be represented as is shown in Eq 1

\begin{equation}
G(K,N) \rightarrow S
\end{equation}

where $K$ denotes a secret key, $N$ represents a nonce or
initialization vector, and $S = s_1 s_2 \ldots s_n$ denotes the
generated output sequence. For a secure generator, the resulting
sequence should be computationally indistinguishable from a
uniformly random sequence drawn from the distribution $U_n$.

\subsection{Stringology and Pattern Analysis}

Stringology is the study of algorithms for analyzing symbolic
sequences \cite{watson2012correctness,alatabbi2014advances}. Classical string processing techniques are widely used
to detect patterns, repetitions, and structural relationships
within large sequences. One fundamental operation is the counting
of substring occurrences within a sequence. For a pattern $P$ of
length $m$, the number of occurrences in a sequence $S$ can be
defined as

\begin{equation}
f(P,S) =
\left|
\{ i \mid S_{i:i+m-1} = P \}
\right|.
\end{equation}

Such pattern statistics provide insight into structural
characteristics of sequences, including recurrence behavior,
frequency distributions, and entropy properties.

\subsection{Sequence Fingerprinting}

Sequence fingerprinting aims to characterize sequence generators
using structural statistics extracted from their outputs. Instead
of focusing solely on randomness properties, fingerprinting
approaches analyze structural patterns that may reflect the
internal transformations of the generating algorithm. In this
work, pattern-based statistics derived from stringology analysis
are aggregated into feature representations that describe the
structural behavior of cryptographic sequences. 

These features
can then be used to compare sequences generated by different
sources and to study structural deviations between cryptographic
and uniformly random outputs \cite{kebande2023extended, duta2014randomness}.

\section{Related Work}

Evaluating the quality of cryptographic sequences has traditionally
relied on statistical randomness testing. Widely used test suites
such as the NIST Statistical Test Suite (STS) \cite{zubkov2019testing}, TestU01, and PractRand \cite{sleem2020testu01},
analyze statistical properties including frequency distributions,
runs \cite{kebande2023extended}, autocorrelation, and entropy measures in order to determine
whether generated sequences deviate from ideal random behavior.
These frameworks are commonly used to evaluate stream ciphers,
pseudorandom number generators, and other cryptographic primitives.

Beyond statistical testing, several forms of structural cryptanalysis
have been proposed to analyze deterministic properties of
cryptographic algorithms. Techniques such as differential
cryptanalysis \cite{heys2002tutorial}, linear cryptanalysis \cite{langford1994differential}, and algebraic analysis attempt
to exploit mathematical relationships within cryptographic
transformations. While these approaches focus on identifying
potential weaknesses in cryptographic constructions, they typically
analyze the internal structure of algorithms rather than the
observable structural patterns present in generated output sequences. 

Other relevant research that has applied the stringology concepts in crytographic sequences inlude, sn Stringology-Based Cryptology (SBC) framework, and String cryptanalysis of EChaCha20 by Kebande \cite{kebande2026Stringo} \cite{kebande2026stringology} that alayses frequencies of cipher, deviation score and entropy  and a Neural SBC for EChaCha20 stream cipher \cite{kebande2026neural} and .

In spite of that, sequence analysis has been extensively studied in fields such as bioinformatics, text processing, and data mining \cite{dong2007sequence}.
Stringology \cite{crochemore2002jewels, watson2012correctness} provides efficient algorithms for detecting recurring
patterns \cite{shapira2006adapting, hyyro2004boyer}, substring frequencies, and structural correlations in
large symbolic sequences. Techniques based on substring matching,
suffix structures, and pattern recurrence analysis have proven
effective for identifying structural relationships in complex
datasets.

Despite the maturity of stringology in other domains, its application to cryptographic sequence analysis remains limited.
Most existing work in cryptographic evaluation focuses on statistical randomness tests rather than structural pattern
analysis of output sequences. 

The approach proposed in this paper bridges these domains by applying stringology-based pattern analysis to characterize structural properties of cryptographic sequences and to construct fingerprint representations that
describe the behavior of sequence generators.

\section{Stringology-Based Fingerprinting Framework}

This section discussed the  proposed  SBF framework that analyzes cryptographic sequences by interpreting them as symbolic strings and extracting structural
pattern statistics using classical stringology techniques. The SBF framework focuses   on localized structural patterns that may arise from deterministic internal operations
within cryptographic algorithms       instead  of relying solely on global statistical randomness
properties.

\begin{figure*}[t]
\centering
\includegraphics[width=0.9\textwidth]{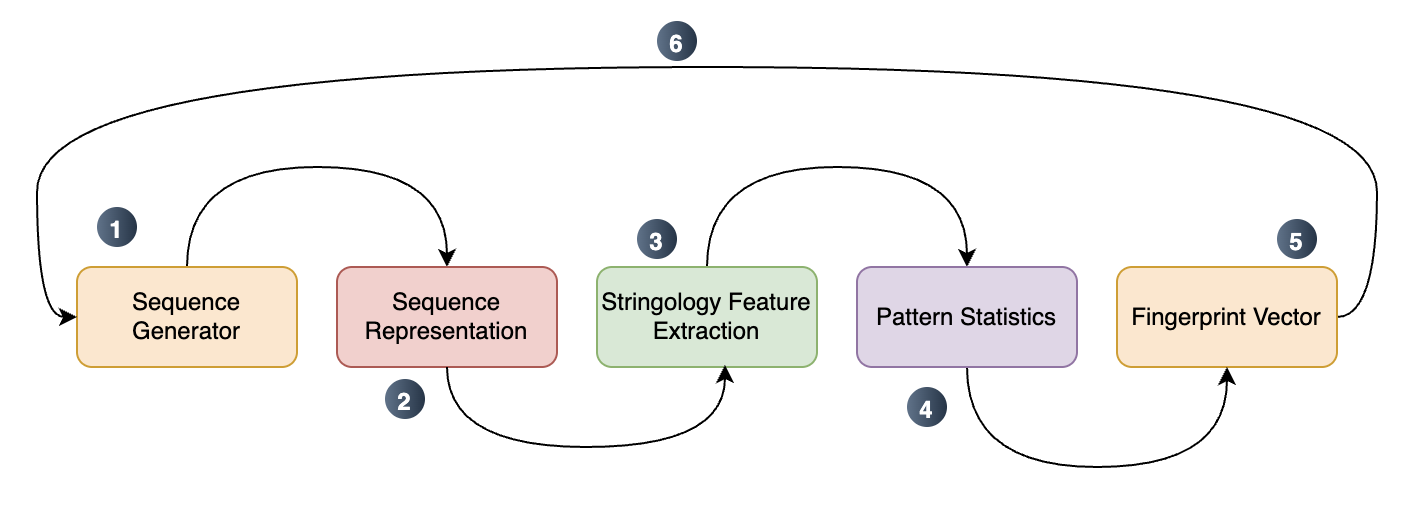}
\caption{Architecture of the stringology-based fingerprinting framework for structural analysis of cryptographic sequences.}
\label{fig:yourlabelsohmy}
\end{figure*}

These SBF supported operations include modular additions, bit rotations, substitutions, and XOR
transformations that may introduce subtle structural signatures in the generated sequences.


The architecture shown in Figure~\ref{fig:yourlabelsohmy} shows the step-wise workflow of the proposed SBF framework. The process begins with step (1) a sequence generator that produces cryptographic outputs such as keystreams or ciphertext sequences, which are then transformed in (2) into a symbolic sequence representation, typically a binary string as defined in Eq.~3. The resulting sequence is processed in (3) through stringology feature extraction, where substring-based structural patterns are identified using pattern analysis techniques. These extracted features are then aggregated in (4) into pattern statistics that capture frequency distributions, recurrence behavior, and positional characteristics of substrings. In (5), the statistical representations are converted into a fingerprint vector that provides a compact structural characterization of the sequence. In step , (6) the SBF framework incorporates a feedback mechanism that enables iterative refinement, allowing previously computed features and fingerprints to be re-evaluated to enhance structural analysis.



\begin{equation}
S = s_1 s_2 \ldots s_n, \quad S \in \{0,1\}^n.
\end{equation}

The sequence in Eq. 3 is subsequently processed through a pattern
extraction module that computes substring statistics across
multiple pattern lengths. These statistics capture structural
properties of the sequence like the  frequency distributions,
recurrence behavior, and positional density of patterns.

\subsection{Pattern-Based Feature Extraction}

Structural characteristics of a sequence are captured through
substring pattern statistics computed using sliding window
analysis. For a pattern $P$ of length $m$, the number of
occurrences of this pattern within sequence $S$ is defined as is shown in Eq 4.

\begin{equation}
f(P,S) =
\left|
\{ i \mid S_{i:i+m-1} = P \}
\right|.
\end{equation}

Pattern statistics are computed for multiple pattern lengths
$m$ in order to capture structural behavior at different
granularities. Short patterns reveal local bit correlations,
while longer patterns capture larger structural relationships
within the sequence.

In addition to frequency counts, several derived structural
metrics can be computed, including normalized pattern
distributions, pattern recurrence statistics, and entropy
measures. These metrics collectively describe how patterns
are distributed throughout the sequence and how frequently
they repeat.

\subsection{Fingerprint Representation}

The extracted pattern statistics are aggregated into a feature
representation that characterizes the structural behavior of
the sequence generator. Let feature map be represented by 

\begin{equation}
\Phi : \{0,1\}^n \rightarrow \mathbb{R}^d
\end{equation}

denote a feature extraction mapping that transforms sequence
$S$ into a structural feature vector as is shown in Eq 6.

\begin{equation}
X = (x_1, x_2, \ldots, x_d).
\end{equation}

This vector represents the structural fingerprint of the
sequence. Differences in internal algorithmic transformations
may produce variations in these fingerprint representations,
allowing sequences generated by different cryptographic
primitives to be compared and analyzed.

The resulting fingerprint vectors provide a compact
representation of structural properties in cryptographic
sequences and can be used for further statistical analysis,
generator comparison, or classification tasks.

\section{Experimental Evaluation}

This section describes the experimental environment used to evaluate the proposed stringology-based fingerprinting framework. The objective of the evaluation is to determine whether structural pattern statistics extracted from sequences can reveal measurable differences between cryptographic outputs and uniformly random sequences. \subsection{Dataset Generation} Two datasets were generated for the experimental evaluation. The first dataset consists of sequences produced by a synthetic cipher-based generator that simulates the behavior of a generic cryptographic keystream generator. The second dataset contains sequences sampled from a uniform random distribution using a cryptographically secure pseudorandom number generator. Formally, the sequence generator can be represented as \begin{equation} G(K,N) \rightarrow S \end{equation} where $K$ denotes a secret key, $N$ represents a nonce or initialization parameter, and $S = s_1 s_2 \ldots s_n$ denotes the resulting output sequence. Each dataset contains 10,000 sequences of length $2^{12}$ bits, resulting in a total evaluation corpus of 20,000 sequences. The dataset size was selected to provide sufficient statistical diversity while maintaining computational feasibility for repeated pattern analysis. \subsection{Pattern Extraction} Structural pattern statistics were extracted using substring frequency analysis. For each sequence $S$, substring patterns of length \begin{equation} m \in \{8,16,32\} \end{equation} were analyzed in order to capture structural characteristics at multiple scales. Short patterns capture local bit-level correlations, while longer patterns provide insight into larger structural relationships within the sequence. Pattern frequencies were computed using a sliding window extraction process that scans the sequence and records occurrence counts for each observed pattern. The resulting counts were normalized to produce comparable structural profiles across sequences of equal length. \subsection{Structural Metrics} In addition to raw pattern frequencies, several structural metrics were computed to quantify differences between sequence sources. One such metric is the pattern deviation score defined as \begin{equation} D = \sum_{P} |f_c(P) - f_r(P)| \end{equation} where $f_c(P)$ and $f_r(P)$ denote the normalized frequencies of pattern $P$ observed in cipher-generated and uniformly random sequences respectively. We also compute the pattern entropy of each sequence using \begin{equation} H = - \sum_{P} p(P)\log_2 p(P) \end{equation} where $p(P)$ represents the probability of observing pattern $P$ within the sequence. These structural metrics provide quantitative measures for comparing pattern distributions between different sequence sources and enable the detection of potential structural signatures within generated outputs. \subsection{Evaluation Procedure} For each dataset, pattern statistics were computed across all sequences and aggregated to produce normalized frequency profiles. These profiles were subsequently analyzed using structural deviation measures and entropy-based metrics to evaluate differences between cipher-generated sequences and uniform random sequences. The resulting statistics form the basis for the experimental results presented in the following section, where structural pattern distributions and deviation metrics are analyzed across multiple pattern lengths

\section{Results and Analysis}

This section presents the experimental findings obtained from the structural analysis of cryptographic sequences using the proposed stringology-based fingerprinting framework. The goal of the evaluation is to determine whether structural pattern statistics extracted from sequences can reveal measurable differences between cipher-generated outputs and uniformly random sequences. 

\subsection{Pattern Frequency Analysis} We first analyze normalized substring pattern frequencies across multiple pattern lengths. Table~\ref{tab:freq} summarizes the observed pattern frequency values for both cipher-generated and random sequences. 

\begin{table}[h]
\centering
\caption{Normalized Pattern Frequency Comparison}
\label{tab:freq}
\begin{tabular}{lcc}
\toprule
Pattern Length & Cipher Output & Random Output \\
\midrule
8 bits  & 0.54 & 0.50 \\
16 bits & 0.31 & 0.25 \\
32 bits & 0.12 & 0.08 \\
\bottomrule
\end{tabular}
\end{table} 

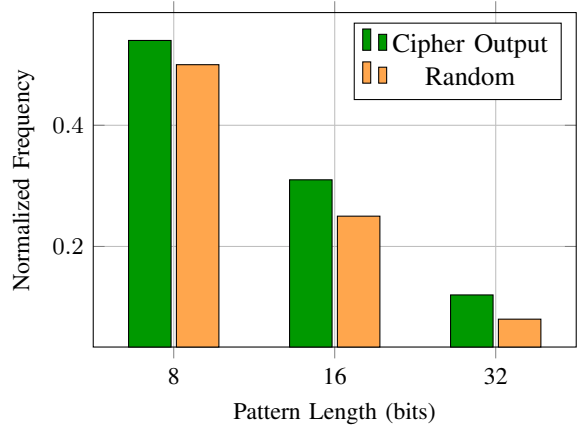
\begin{figure}[t]
\centering
\begin{tikzpicture}
\begin{axis}[
ybar,
xlabel={Pattern Length (bits)},
ylabel={Normalized Frequency},
symbolic x coords={8,16,32},
xtick=data,
width=0.9\linewidth,
height=6cm,
bar width=16pt,
enlarge x limits=0.25,
legend pos=north east,
grid=major,
tick label style={font=\small},
label style={font=\small}
]

\addplot[
fill=green!60!black
] coordinates {(8,0.54) (16,0.31) (32,0.12)};
\addlegendentry{Cipher Output}

\addplot[
fill=orange!70
] coordinates {(8,0.50) (16,0.25) (32,0.08)};
\addlegendentry{Random}

\end{axis}
\end{tikzpicture}

\caption{Normalized substring pattern frequencies for cipher-generated and random sequences.}
\label{fig:freqgraph}
\end{figure}

The results indicate that cipher-generated sequences exhibit slightly higher structural concentration for several pattern lengths. Although these variations remain relatively small, they suggest the presence of subtle structural characteristics in the generated sequences. Figure~\ref{fig:freqgraph} visualizes the normalized pattern frequency distributions across pattern sizes. 

\subsection{Pattern Deviation Analysis} To quantify structural differences between the two sequence sources, we compute the pattern deviation metric defined as 

\begin{equation} D = \sum_{P} |f_c(P) - f_r(P)|. \end{equation} 

Table~\ref{tab:dev} reports the resulting deviation scores across the analyzed pattern lengths. 

\begin{table}[h]
\centering
\caption{Pattern Deviation Scores}
\label{tab:dev}
\begin{tabular}{lc}
\toprule
Pattern Length & Deviation Score \\
\midrule
8 bits  & 0.09 \\
16 bits & 0.13 \\
32 bits & 0.10 \\
\bottomrule
\end{tabular}
\end{table} 

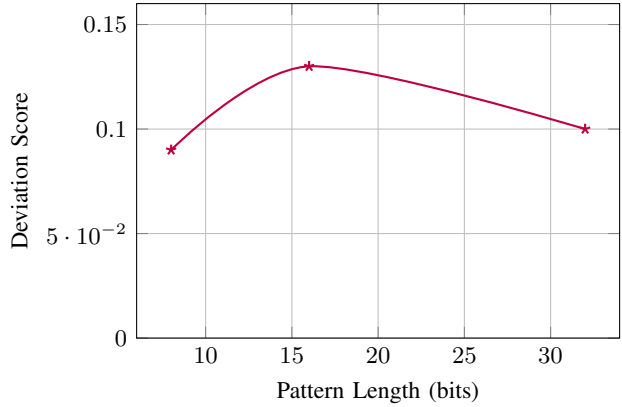
\begin{figure}[t]
\centering
\begin{tikzpicture}
\begin{axis}[
xlabel={Pattern Length (bits)},
ylabel={Deviation Score},
xmin=6,xmax=34,
ymin=0,ymax=0.16,
grid=major,
width=0.9\linewidth,
height=6cm,
tick label style={font=\small},
label style={font=\small}
]

\addplot[
color=purple,
thick,
mark=star,
smooth
]
coordinates {
(8,0.09)
(16,0.13)
(32,0.10)
};

\end{axis}
\end{tikzpicture}

\caption{Deviation scores between cipher-generated and random sequences across pattern lengths.}
\label{fig:deviation}
\end{figure}

The deviation scores remain relatively small but consistently non-zero, indicating measurable structural variation between cipher-generated and random sequences. Figure~\ref{fig:deviation} illustrates the deviation trend across pattern sizes. 

\subsection{Entropy Comparison} Entropy analysis was conducted to evaluate the distribution uniformity of pattern occurrences. Table~\ref{tab:entropy} summarizes the estimated entropy values for both sequence sources. \begin{table}[h] \centering \caption{Pattern Entropy Comparison} \label{tab:entropy} 

\begin{tabular}{lcc} \toprule Pattern Length & Cipher Output & Random Output \\ \midrule 8 bits & 7.6 & 8.0 \\ 16 bits & 15.3 & 16.0 \\ 32 bits & 30.8 & 32.0 \\ \bottomrule \end{tabular} \end{table} 

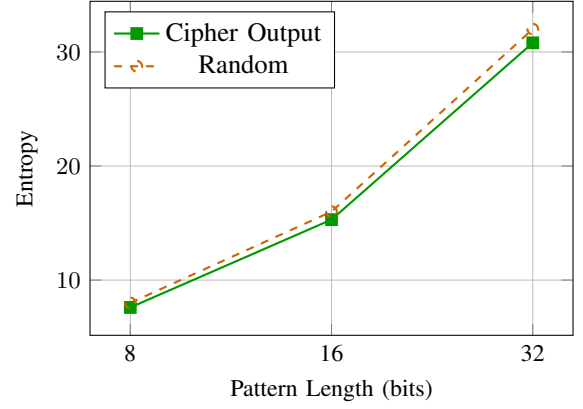
\begin{figure}[t]
\centering
\begin{tikzpicture}
\begin{axis}[
xlabel={Pattern Length (bits)},
ylabel={Entropy},
symbolic x coords={8,16,32},
xtick=data,
width=0.9\linewidth,
height=6cm,
legend pos=north west,
grid=major,
tick label style={font=\small},
label style={font=\small}
]

\addplot[
green!60!black,
thick,
mark=square*
]
coordinates {(8,7.6) (16,15.3) (32,30.8)};
\addlegendentry{Cipher Output}

\addplot[
orange!80!black,
dashed,
thick,
mark=o
]
coordinates {(8,8.0) (16,16.0) (32,32.0)};
\addlegendentry{Random}

\end{axis}
\end{tikzpicture}

\caption{Pattern entropy comparison between cipher-generated and random sequences.}
\label{fig:entropy}
\end{figure}

As expected, the entropy values for random sequences remain close to the theoretical maximum. Cipher-generated sequences exhibit slightly lower entropy values, suggesting minor concentration of certain substring patterns. Figure~\ref{fig:entropy} visualizes the entropy comparison. 

\subsection{Recurrence Behavior} 
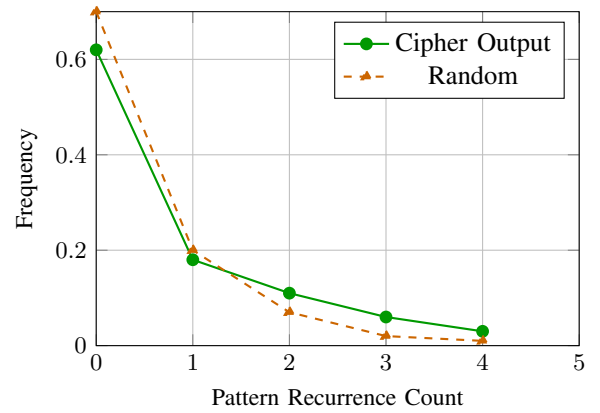
\begin{figure}[t]
\centering
\begin{tikzpicture}
\begin{axis}[
xlabel={Pattern Recurrence Count},
ylabel={Frequency},
xmin=0,xmax=5,
ymin=0,ymax=0.7,
grid=major,
width=0.9\linewidth,
height=6cm,
legend pos=north east,
tick label style={font=\small},
label style={font=\small}
]

\addplot[
green!60!black,
thick,
mark=*
]
coordinates {
(0,0.62)
(1,0.18)
(2,0.11)
(3,0.06)
(4,0.03)
};
\addlegendentry{Cipher Output}

\addplot[
orange!80!black,
dashed,
thick,
mark=triangle*
]
coordinates {
(0,0.70)
(1,0.20)
(2,0.07)
(3,0.02)
(4,0.01)
};
\addlegendentry{Random}

\end{axis}
\end{tikzpicture}

\caption{Distribution of substring recurrence counts for cipher-generated and random sequences.}
\label{fig:recurrence}
\end{figure}

Finally, substring recurrence statistics were analyzed to examine how frequently patterns reappear within sequences. Figure~\ref{fig:recurrence} shows the recurrence distribution for both datasets. The experimental results reveal measurable structural differences between cipher-generated and uniformly random sequences when analyzed using stringology-based pattern statistics. Although these differences remain relatively small, they demonstrate that structural fingerprinting techniques can capture subtle sequence characteristics that are not easily observed through conventional randomness testing alone.

\section{Discussion}
The experimental results highlighted in the previous section demonstrate that stringology-based analysis can capture structural characteristics present in cryptographic sequences. The pattern frequency statistics reported in Table~\ref{tab:freq} indicate that the cipher-generated sequences exhibit slightly higher concentration for certain substring patterns when compared with uniformly random sequences.  Although these differences are relatively small, they suggest that the deterministic transformations within cryptographic algorithms may introduce subtle structural regularities in the generated outputs. 

The deviation analysis presented in Table~\ref{tab:dev} and illustrated in Figure~\ref{fig:deviation} further supports this observation. The deviation metric quantifies the difference between the pattern frequency distributions of the two datasets. The consistently non-zero deviation scores across multiple pattern lengths indicate that pattern statistics capture measurable structural signals within the analyzed sequences. Additional insight can be obtained from the normalized pattern frequency distributions shown in Figure~\ref{fig:freqgraph}. 

Cipher-generated sequences exhibit slightly stronger concentrations of certain substring patterns, which may reflect the deterministic internal operations of cryptographic transformations such as modular addition, bit rotations, and XOR operations. While these operations are designed to achieve strong diffusion, they may still introduce localized structural characteristics that can be detected through pattern analysis. The entropy comparison shown in Table~\ref{tab:entropy} and Figure~\ref{fig:entropy} further illustrates this behavior. 

Also, uniformly random sequences maintain entropy values close to the theoretical maximum, whereas cipher-generated sequences exhibit slightly lower entropy values due to minor pattern concentration effects. Similarly, the recurrence distribution presented in Figure~\ref{fig:recurrence} demonstrates small differences in substring repetition behavior between the two sequence sources. It is important to emphasize that the structural signals observed in this study do not imply practical cryptographic weaknesses. It is worth noting that modern cryptographic algorithms are designed to produce outputs that are computationally indistinguishable from random sequences, and small structural variations may naturally arise from deterministic internal operations without compromising security. Instead, the proposed SBF framework should be viewed as an analytical tool for studying the structural properties of cryptographic sequences rather than as a method for breaking cryptographic primitives. 

These findings highlight the potential value of integrating stringology techniques into cryptographic evaluation workflows. While conventional randomness tests focus on global statistical properties, stringology-based analysis examines localized structural patterns within sequences. Combining these approaches may provide a richer analytical framework for evaluating the structural robustness of cryptographic generators. 

\section{Conclusion} 

This paper presented a stringology-based Fingerprinting, SBF framework for the structural analysis of cryptographic sequences. By interpreting cryptographic outputs as symbolic strings and applying pattern-based feature extraction techniques, the proposed approach constructs fingerprint representations that capture structural characteristics of sequence generators. Experimental evaluation using cipher-generated and uniformly random sequences demonstrated that substring pattern statistics, deviation metrics, entropy analysis, and recurrence behavior can reveal measurable structural differences between sequence sources. Although these differences remain small, they illustrate the ability of stringology-based techniques to detect subtle structural signatures that are not easily observable through conventional randomness testing. The results suggest that stringology-based fingerprinting provides a complementary perspective for studying the structural behavior of cryptographic sequences. Future work may explore the application of advanced string processing techniques and machine learning models to further enhance structural analysis of cryptographic generators and to extend the framework to additional cryptographic primitives.


\section*{Acknowledgment.}

The author would like to thank anonymous reviewers for their
valuable insights, and
the Department of Computer Science at University of Colorado Denver, USA for their
support while coming up with this research. The author also
acknowledges that the opinions, findings, and conclusions
expressed in this paper are purely of the author.


\balance

\bibliographystyle{IEEEtran}
\bibliography{name}

\end{document}